\documentclass[aps,twocolumn,pra]{revtex4-1}
\usepackage{graphics}
\usepackage{dcolumn}
\usepackage{bm}
\usepackage{amsmath}
\usepackage{hyperref}
\hypersetup{colorlinks=true, urlcolor=blue}
\usepackage{color}           

\usepackage{newlfont}
\usepackage{amssymb}
\usepackage{amsfonts}
\usepackage{amsmath}
\usepackage{graphicx}
\usepackage{bm}

\def \be{\begin{equation}}
\def \ee{ \end{equation} }

\begin{document}
\renewcommand*{\DefineNamedColor}[4]{%
   \textcolor[named]{#2}{\rule{7mm}{7mm}}\quad
  \texttt{#2}\strut\\}

\definecolor{red}{rgb}{1,0,0}
\title{Dual quantum-correlation paradigms exhibit opposite statistical-mechanical properties}
\author{R. Prabhu, Aditi Sen(De), and Ujjwal Sen}

\affiliation{Harish-Chandra Research Institute, Chhatnag Road, Jhunsi, Allahabad 211 019, India}

\begin{abstract}
We report opposite statistical mechanical behaviors of the two major paradigms in which quantum correlation measures are defined, viz., the entanglement-separability paradigm and the information-theoretic one. 
We show this by considering the ergodic properties of such quantum correlation measures in transverse quantum $XY$ spin-$\frac{1}{2}$ systems in low dimensions. While entanglement measures are ergodic in such models, 
the quantum correlation measures defined from an information-theoretic perspective can be nonergodic. 
\end{abstract}

\maketitle

\section{Introduction}
Quantum correlations -- being one of the most important physical resources in quantum information science \cite{Nielsen00} -- 
has led to the development of technologies based on quantum principles. 
There has been a continuous effort towards characterizing, quantifying, and utilizing quantum correlations for nearly  
the last two decades. Understanding quantum correlations theoretically and their robust experimental generation remain important goals, although 
significant advances have been achieved in recent years on both fronts.

Of the different measures that have been put forward to quantify and conceptualize quantum correlations, there are broadly two families. 
The first contains the ones that are defined within the 
entanglement-separability paradigm \cite{Horodecki09}, and they play a key role in describing quantum information phenomena. 
Indeed, entanglement is the resource that 
drives the potentially revolutionizing applications in quantum information, including secure cryptography \cite{Gisin02}, faster computers \cite{qcompureview}, 
and better communication channels \cite{qcommreview}. 
Recently, there has also been a considerable interest to 
study entanglement in quantum many-body systems \cite{Lewenstein07,Amico08,Bloch08}.


Quantum correlations have also been conceptualized from an information-theoretical point of view -- these forming the second family. 
Such correlations have the potential to be useful in understanding phenomena 
in which entanglement
may not be necessary 
\cite{nlweetc, noentresource,noentresourceexp}. 
Important examples in this family are 
%
%
quantum discord (QD) \cite{discord1} and quantum work-deficit \cite{deficit}. 
Quite naturally, QD and quantum work-deficit 
have also been used to study many-body phenomena \cite{rd08}.

Investigations of ergodicity of  a physical quantity in a many-body system plays an important role 
in understanding statistical-mechanical properties of that system.  
The validity of a statistical-mechanical 
description of a quantity 
that characterizes a physical system depends on the behavior of that quantity in time-evolved states in comparison to 
that in the corresponding equilibrium states. 
A physical quantity is said to
be ergodic if the time average of the quantity matches with its ensemble average. 
In this paper, we consider the status of ergodicity of quantum correlation measures in the quantum anisotropic transverse \(XY\) models, in 
low-dimensional spin-$\frac{1}{2}$ lattices.
The question of ergodicity of magnetization and classical correlations were considered in 
the quantum $XY$ chain 
in Refs. \cite{Mazur69, Perk77, Barouch1075,Barouch786, amaderergo1, amaderergo2}. 
Advances in the science of quantum information leads us to consider the question of ergodicity also for \emph{quantum} correlation measures.
Here we show that quantum correlation concepts defined from independent perspectives can exhibit opposing statistical-mechanical behaviors. 
We compare the ergodicity behaviors of quantum correlations defined within the entanglement-separability paradigm, 
such as
concurrence \cite{concurrence} and logarithmic negativity \cite{VidalWerner}, with those in the information-theoretic one, 
in the quantum $XY$ spin systems in a transverse field in one dimension, ladder, and two dimensions.
We find that while entanglement measures 
display
ergodicity irrespective of the initial field strengths, quantum discord and quantum work-deficit are ergodic for both low and high transverse 
magnetic fields and nonergodic for moderate 
ones. 
For low fields, we prove  ergodicity quite generally: Arbitrary quantum correlation measures, irrespective of whether defined in the entanglement-separability paradigm 
or by using information-theoretic concepts, are ergodic, for any interacting system in a non-commuting step function field, in any dimension. 
Moreover, we analytically show that for high transverse magnetic fields, all quantum correlation measures are ergodic in the infinite quantum transverse $XY$ chain
 for arbitrary anisotropy and arbitrary initial temperature. Exact diagonalization techniques are used to deal with high fields in higher dimensional systems. For moderate
field strength, the Jordan-Wigner transformation is utilized in the case of an infinite one-dimensional (1D) system, while exact diagonalization is used in higher dimensions.

The paper is organized as follows. In Sec. \ref{sec:XYmodel}, we give a brief description of the model that 
we consider, and give simplified forms for the single- and two-site reduced density matrices of the model.
 In Sec. \ref{sec:QCmeas}, we define the logarithmic negativity and concurrence as measures of quantum entanglement and
also define the information-theoretic measures, viz., quantum discord and quantum work-deficit. 
The notion of ergodicity is formally presented in Sec. IV, and the same section also contains the general analytical ergodicity result for low fields. 
We study the ergodicity of the different quantum correlation measures for the infinite quantum spin chain in Sec. \ref{sec:inf},
where we also present the analytical proposition for ergodicity at high fields.
The  ergodicity question for 1D finite spin chains, of moderate sizes, is considered in Sec. \ref{sec:finite1d}, and
we find that the answer is qualitatively the same as in the infinite case. 
Sections \ref{sec:ladder} and
\ref{sec:2d} address the question of ergodicity, for the ladder and 2D quantum $XY$ spin models respectively.
Finally, we summarize our results in Sec. \ref{summary}.

\section{The \(XY\) model in a transverse field}
\label{sec:XYmodel}

\subsection{Description of the model}

An interacting  Hamiltonian of spin-$\frac{1}{2}$ particles on a lattice, with an external transverse magnetic field, is given by 
$H = H_{\mathrm{int}} - h(t) H_{\mathrm{mag}},$
where $h(t)$ is a  time-dependent transverse magnetic field. To ensure that the field part of Hamiltonian has a nontrivial effect on the system dynamics, 
one should have
$[H_{\mathrm{int}}, H_{\mathrm{mag}}]\neq 0$. 
A simple way to obtain this is by choosing the field part of the Hamiltonian as
$H_{\mathrm{mag}} = \sum_i S_i^z,$
and the interaction part as
$H_{\mathrm{int}} = J \sum_{\langle ij \rangle} \left[(1 + \gamma) S_i^x S_{j}^x + (1 - \gamma) S_i^y S_{j}^y\right],$
with $\gamma \ne 0$, where $J$ is the coupling constant, 
$\gamma$ is the anisotropy measure, $S_i^x$, $S_i^y$, and $S_i^z$ are one-half of the Pauli spin matrices at the 
$i^{\mathrm{th}}$ lattice site, 
and $\langle ij\rangle$ indicates that the corresponding sum runs over 
nearest-neighbor pairs of the lattice. Periodic boundary conditions are considered in all 
the models dealt with in this paper. The total Hamiltonian studied, therefore, takes the form 
\begin{equation}
\label{asolH}
H(t) = J \sum_{\langle ij \rangle} \left[(1 + \gamma) S_i^x S_{j}^x + (1 - \gamma) S_i^y S_{j}^y\right] - h(t) \sum_i S_i^z.
\end{equation}
This model is known as the anisotropic (quantum) $XY$ model in a transverse field. 
The time-dependent magnetic field \(h(t)\) is chosen as
\begin{equation}
h(t) = \Big\{\begin{array}{c} a,  \quad t \leq 0, \\ 0, \quad t  >  0,  \end{array}
\label{Eq:transfield}
\end{equation}
with \(a \ne 0\).
Set \(\tilde{h}(t) = h(t)/J\), \(\tilde{a} = a/J\). 
Although we carry out our investigations by considering a vanishing magnetic field for \(t>0\), 
the results remain qualitatively similar even for non-vanishing applied field strengths at nonzero time, i.e. for $h(t) \neq 0$ for $t>0$. 




\subsection{Single- and two-site reduced density matrices}

 In order to evaluate the quantities that are of interest in this paper, 
we need to find the single-site and two-site reduced density matrices of the equilibrium and evolved states of the system.

A general single-site density matrix is given by
\begin{equation*}
\rho^{\textit 1}=\frac{1}{2}\left(I+\vec M\cdot\vec \sigma \right),
\label{eq:}
\end{equation*}
where $I$ denotes the identity on the qubit Hilbert space and
$\vec M=\mathrm{tr}[\rho^{\textit 1}\vec \sigma]$ is the magnetization vector, where \(\vec \sigma = \left( \sigma^x, \sigma^y, \sigma^z \right)\)
are the Pauli matrices.
Let us first consider the single-site density matrix for 
the equilibrium state, \(\rho_{eq}^\beta(t)\), at temperature $T$, where $\beta=\frac{1}{k_BT}$, with $k_B$ being the Boltzmann constant. Due to symmetry, all the single-site density matrices of the equilibrium states are equal.
We will denote them by \(\rho_{eq}^{\textit{1}}(t)\) (hiding the prefix \(\beta\)).
Now \(\rho_{eq}^{\textit{1}*}(t) = \rho_{eq}^{\textit{1}}(t)\), where the complex conjugation 
is taken in the computational basis, which (for each site) is the eigenbasis of the Pauli matrix \(\sigma^z\).
Therefore $$M_{eq}^y(t) = \mathrm{tr} [\sigma^y \rho_{eq}^{\textit{1}}(t)]=0.$$
Moreover the Hamiltonian \(H(t)\) has the global  phase flip symmetry (\([H,\Pi_i \sigma^z_i]=0\)), from which it follows that 
$$M_{eq}^x(t) = \mathrm{tr} [\sigma^x \rho_{eq}^{\textit{1}}(t)]=0.$$ 
Consequently, the single-site density matrix of the equilibrium state reduces to
$$\rho_{eq}^{\textit{1}}(t) = \frac{1}{2}\left(I + M_{eq}^z(t)\sigma^z\right).$$
The evolved state does not necessarily have the property of being equal to its complex conjugation,
and consideration of the global phase flip symmetry is complicated 
by the fact that the Hamiltonian is explicitly dependent on time.
However, using the Wick's theorem, as in \cite{LSM61,Barouch1075,Barouch786},
the single-site density of the time-evolved state again turns out to be of the form
$$\rho^{\textit 1}(t) = \frac{1}{2}\left(I + M^z(t)\sigma^z\right).$$
So, the single-site (transverse) magnetization of the equilibrium state is 
\begin{equation}
M_{eq}^z(t) = \mathrm{tr}[\sigma^z \rho_{eq}^{\textit{1}}(t)], \nonumber
\end{equation}
while that for the evolved state is
\begin{equation}
M^z(t) = \mathrm{tr}[\sigma^z \rho^{\textit{1}}(t)]. \nonumber
\end{equation}

The general two-site density matrix for the equilibrium as well as the time-evolved state can therefore be written as
\begin{equation}
\rho^{\textit {12}}=\frac{1}{4}\big[I\otimes I+M^z(\sigma^z\otimes I+I \otimes \sigma^z)+\sum_{i,j=x,y,z}T^{ij}(\sigma^i\otimes\sigma^j)\big],\nonumber
\end{equation}
where 
$T^{ij}=\mathrm{tr}[\rho^{\textit {12}}(\sigma^i\otimes\sigma^j)]$ represents two-site correlation functions. Once again by using Wick's theorem, we 
can show that the \(yz\) and \(xz\) correlations will vanish for the evolved state, and for large time, also the \(xy\) correlation vanishes. 
For the equilibrium state, only the diagonal correlations, \(T^{xx}\), \(T^{yy}\), and 
\(T^{zz}\), remain.

\section{Quantum correlation measures}
\label{sec:QCmeas}

\subsection{Measures in entanglement-separability paradigm}

In this subsection, we will define two quantum correlation measures within the entanglement-separability paradigm. Both have the virtue of being efficiently computable for the case of two-qubit states, the latter being the quantum states at the focus of our study.

\subsubsection{Concurrence}
\label{concurrence}

The concept of concurrence came from the definition of entanglement of formation and is proposed to quantify entanglement
in two-qubit (possibly mixed) states \cite{concurrence}. 
The entanglement of formation of a bipartite quantum state is the amount of singlets, \(\frac{1}{\sqrt{2}}(|01\rangle - |10\rangle)\),  that are required to prepare 
the state by local quantum operations and classical communication (LOCC). Here \(|0\rangle\) and \(|1\rangle\) form an
orthonormal qubit basis.
For two-qubit states, \(\rho_{AB}\), there exists a closed form of the entanglement of formation in terms of 
the concurrence, which is given by  
$C(\rho_{AB})=\mbox{max}\{0,\lambda_1-\lambda_2-\lambda_3-\lambda_4\}$, where the
$\lambda_i$'s are the square roots of the eigenvalues of $\rho_{AB}\tilde{\rho}_{AB}$ in decreasing order and 
$\tilde{\rho}_{AB}=(\sigma_y\otimes\sigma_y)\rho_{AB}^*(\sigma_y\otimes\sigma_y)$, with the complex conjugation being taken
in the computational  basis.

\subsubsection{Logarithmic negativity}
\label{entanglement}

Another measure of entanglement considered here is the 
logarithmic negativity (LN) \cite{VidalWerner}.
The negativity, \(N(\rho_{AB})\), of a bipartite state \(\rho_{AB}\) is defined as the absolute value of the sum of the negative eigenvalues of \(\rho_{AB}^{T_{A}}\),
 where \(\rho_{AB}^{T_{A}}\) denotes the partial transpose of \(\rho_{AB}\) with respect to the \(A\)-part \cite{Peres_Horodecki}. The logarithmic negativity is defined as
\begin{equation}
  E_{N}(\rho_{AB}) = \log_2 [2 N(\rho_{AB}) + 1].
\label{eq:LN}
\end{equation}
For two-qubit states, 
\(\rho_{AB}^{T_{A}}\) has at most one negative eigenvalue \cite{Anna-ek}. Moreover, for two-qubit states, 
a positive LN implies that the state is entangled and distillable \cite{Peres_Horodecki, Horodecki_distillable}, while \(E_{N} =0\) 
implies that the state is separable \cite{Peres_Horodecki}.

\subsection{Information-theoretic measures}

In this subsection, we will define two measures of quantum correlation defined from an information-theoretic perspective. In contrast to the entanglement measures 
defined in the preceding subsection, these are not computable in closed form. However, they can be efficiently computed, through numerical 
simulations, for two-qubit systems.

\subsubsection{Quantum discord}
\label{discord}

Quantum discord is defined as the difference between two quantum information-theoretic quantities, whose classical counterparts are 
equivalent expressions for the classical mutual information
\cite{discord1}: 
\begin{equation}
\label{eq:discord}
Q(\rho_{AB})= {\cal I}(\rho_{AB}) - {\cal J}(\rho_{AB}).
\end{equation}
The total correlation, \({\cal I}(\rho_{AB})\), of a bipartite state \(\rho_{AB}\) is given by \cite{qmi} (see also \cite{Cerf, GROIS})
\begin{equation}
\label{qmi}
\mathcal{I}(\rho_{AB})= S(\rho_A)+ S(\rho_B)- S(\rho_{AB}),
\end{equation}
where $S(\varrho)= - \mbox{tr} (\varrho \log_2 \varrho)$ is the von Neumann entropy of the quantum state \(\varrho\), and 
 \(\rho_A\) and \(\rho_B\) are the reduced density matrices of  \(\rho_{AB}\).
On the other hand, \({\cal J}(\rho_{AB})\) can be interpreted as the amount of classical correlation in \(\rho_{AB}\), and is defined as 
\begin{equation}
\label{eq:classical}
 {\cal J}(\rho_{AB}) = S(\rho_A) - S(\rho_{A|B}). 
\end{equation}
Here
\begin{equation}
 S(\rho_{A|B}) = \min_{\{B_i\}} \sum_i p_i S(\rho_{A|i}),
\end{equation}
is the conditional entropy of \(\rho_{AB}\), conditioned on a measurement performed by \(B\) with a rank-one projection-valued measurement \(\{B_i\}\),
producing the states  
\(\rho_{A|i} = \frac{1}{p_i} \mbox{tr}_B[(\mathbb{I}_A \otimes B_i) \rho (\mathbb{I}_A \otimes B_i)]\), 
with probability \(p_i = \mbox{tr}_{AB}[(\mathbb{I}_A \otimes B_i) \rho (\mathbb{I}_A \otimes B_i)]\).
\(\mathbb{I}\) is the identity operator on the Hilbert space of \(A\).

\subsubsection{Quantum work-deficit}
\label{sec:workdeficit}

Another important information-theoretic measure of quantum correlation is quantum work-deficit \cite{deficit}, 
which is defined for a bipartite quantum state 
as the difference between the amount of pure states that can be extracted under global operations and pure product states that 
can be extracted under 
local operations, in closed systems for which addition of the corresponding pure states are not allowed.

The allowed class 
of global operations, known as ``closed global operations'' (CGO) are any sequence of (G1) unitary operations, and 
(G2) dephasing of a given state $\rho_{AB}$ by using a 
set of projectors $\{P_i\}$, i.e., $\rho \rightarrow \sum_i P_i \rho_{AB} P_i$,  
where $P_iP_j = \delta_{ij} P_i$, $\sum_i P_i = \mathbb{I}$, with 
$\mathbb{I}$ being the identity operator on the Hilbert space ${\cal H}$ on which $\rho_{AB}$ is defined. 
The number of pure qubits that can be extracted from $\rho_{AB}$ by CGO is 
\[I_G (\rho_{AB})= N - S(\rho_{AB}),\]
 where $N = \log_2 (\dim {\cal H})$.
The number of qubits that can be extracted from a bipartite quantum state $\rho_{AB}$ under 
``closed local operations and classical communication''(CLOCC), which consists of local unitary, 
local dephasing, and sending dephased state from one party to another, is defined as
\begin{equation}
I_L(\rho_{AB}) = N - \inf_{\Lambda \in CLOCC} [S(\rho{'}_A) + S(\rho{'}_B)],
\end{equation}
where $S(\rho{'}_A) = \mbox{tr}_B (\Lambda (\rho_{AB}))$  and $S(\rho{'}_B) = \mbox{tr}_A (\Lambda (\rho_{AB}))$. 

Quantum work-deficit 
is  defined as the difference between work, $I_G (\rho_{AB})$,  extracted by CGO,  and that by CLOCC, $I_L (\rho_{AB})$:
\begin{equation}
 \Delta(\rho_{AB}) = I_G(\rho_{AB}) - I_L(\rho_{AB}).
\end{equation}
The quantity is not efficiently computable for arbitrary states, and therefore we restrict our attention to CLOCC consisting of measurement at a 
single party only. This restricted work-deficit is equivalent to quantum discord for bipartite states having maximally mixed marginals.


\section {Ergodicity}

In this section, we will formally describe the notion of ergodicity of a physical parameter, and  will subsequently 
show that any quantum correlation measure is ergodic when the initial
magnetic field is very low, irrespective of the other details of the Hamiltonian, and its dimension.
  
To study the ergodicity of a physical quantity ${\cal P}$ in a given physical system, we evolve the system 
from an initial canonical equilibrium state 
at a given temperature $T$, with the physical system being described by a Hamiltonian \(H= H_{int} - h(t) H_{mag}\), where \(h(t)\) involves an initial kick to the system.
 The behavior of the time-averaged value, at large times,
of ${\cal P}$, in the time-evolved state of the system is denoted by \({\cal P}^\infty(T,\tilde{h}(t), \tilde{\gamma})\),
 where \(\tilde{\gamma}\) consists of the aggregate of all relevant parameters in the system, other than \(T\) and \(\tilde{h}(t)\).
This time-averaged quantity is then compared to 
that of the canonical equilibrium state at large times and at temperature \(T{'}\), denoted by 
\({\cal P}^{\mathrm{can}}(T{'}, \tilde{h}(t=\infty), \tilde{\gamma})\).
%
%
The physical quantity \({\cal P}\) will be called nonergodic, if for a given \(T\),
\begin{equation}
\label{raat-duto-egaro}
{\cal P}^\infty(T,\tilde{h}(t), \tilde{\gamma}) \ne {\cal P}^{\mathrm{can}}(T{'},\tilde{h}(t=\infty), \tilde{\gamma}) \quad \forall \, T{'},
\end{equation}
where \(T{'}\) is chosen in a physically relevant range.
Otherwise, it will be termed ergodic.
 We term a physical quantity as strongly nonergodic if the relation (\ref{raat-duto-egaro}) holds for all temperatures \(T{'}\). 
Note here that the question of ergodicity in Ref. \cite{amaderergo1} was addressed by considering the effective temperature for a dynamical (time-evolved) state of the $XY$ spin chain,
which was determined by the condition that the equilibrium state must be on the same energy surface as that of the evolved state.
Here we remove the energy constraint in  investigating the ergodicity properties of different physical quantities. 
In other words, the temperature $T{'}$ is now not fixed by the energy (or any other) constraint,
but is allowed to be within a certain 
relevant range of the initial temperature $T$. The length of this range can depend on several uncontrolled parameters that may be active in an experimental realization of the theoretical model, including 
decoherence effects, disorder-induced effects, etc.


The above definition of ergodicity is ``microcanonical-like'', in the sense that for a particular physical quantity to be ergodic, 
one must have equality in Eq. (\ref{raat-duto-egaro}). Such a definition is however of limited  realistic applicability, as fluctuations
 of the corresponding physical values (on both sides of Eq. (\ref{raat-duto-egaro})) are disregarded. It is physically more meaningful to call a physical quantity as ergodic if 
\begin{eqnarray}
\label{dupur-tinte-bajte-dosh}
 {\cal P}^\infty(T,\tilde{h}(t), \tilde{\gamma})
 \in \phantom{phaltu phaltu phaltu phaltu phaltu } \nonumber \\
({\cal P}^{\mathrm{can}}(T{'},\tilde{h}(t=\infty), \tilde{\gamma})  - \epsilon , 
{\cal P}^{\mathrm{can}}(T{'},\tilde{h}(t=\infty), \tilde{\gamma}) + \epsilon), \nonumber\\
\end{eqnarray}
for some \(T{'}\), and where the choice of \(\epsilon\) is to be elaborated below. 
A physical quantity which is ergodic according to Eq. (\ref{dupur-tinte-bajte-dosh}) will be termed as ``canonically ergodic''.
 A quantity that is ergodic (i.e. it satisfies Eq. (\ref{raat-duto-egaro}) with an equality, for some \(T{'}\)) is of course canonically so. 
The choice of \(\epsilon\) will depend on the physical system under consideration, and the physical quantity that we are studying.
 Typically, it will be of the same magnitude as the error in evaluating that quantity in an experimental implementation of the system. Where such an error is not apparent, 
one may choose \(\epsilon\) as the standard deviation of the physical quantity in the ensemble that is being considered.

\noindent\textbf{Proposition 1.} Consider the Hamiltonian \(H =   H_{int} - h(t) H_{mag}\), where \([H_{int},  H_{mag}] \neq 0\). The time-dependent ``magnetic field'', \(h(t)\),
 is a step function and is of the form \(h(t) = a, t\leq 0\) and \(h(t) =0, t>0\). 
  For low initial field \(a\), i.e. 
in the neighborhood of \(a \approx 0\), all quantum correlation measures are ergodic,
 in any dimension.  

\noindent Proof.
If we choose \(a \) to be zero which is the same as the magnetic field for \(t>0\),
then there will be no evolution in the system and 
hence all the quantum correlations of the evolved state will be the same as that of the initial thermal equilibrium state.  
In the vicinity of \(a \approx 0\), and with the initial temperature \(T\),
it is always possible to find a temperature \(T^{'} \approx T\) for which 
all quantum correlation measures of the evolved state will be same as that of the thermal
equilibrium state at \(T^{'}\). Here we assume that \(H_{int}\) and \(H_{mag}\) do not depend on time.\hfill \(\square\)

Although we state the proposition for low fields, the theorem holds for any value of \(a\) which is close to a constant \(h(t)\) for \(t> 0\).

\section{Infinite Quantum $XY$ spin chain in a transverse field}
\label{sec:inf}
We now study and  compare the statistical-mechanical behavior of quantum correlation concepts defined in the two qualitatively different paradigms, viz.,
the entanglement-separability and the information-theoretic ones, as described in Sec. \ref{sec:QCmeas}, for the infinite  1D \(XY\) model (see Eq. (\ref{asolH})).
From Prop. 1, we obtain that for low fields, the measures from both the paradigms will exhibit ergodicity. 
Here we show, analytically, that for high fields, they are also ergodic. 
In contrast, moderate magnetic fields produce opposite effects in the two paradigms, as we will show later.

\noindent\textbf{Proposition 2.} For high fields i.e. for \(\tilde{a} \rightarrow \infty\), any quantum correlation measure is ergodic for the 1D transverse $XY$ model for arbitrary
initial temperature  and the anisotropy.

\noindent Proof. The 1D transverse $XY$ model lends itself to exact diagonalization  \cite{LSM61,Barouch1075,Barouch786}. We perform the corresponding
Jordan-Wigner and  Fourier transformations successively and find that 
the transverse magnetization in the time-evolved state at long times, considering that the initial state is the canonical equilibrium state at temperature \(T\), is given by
\begin{eqnarray}
M^z &=& -\frac{1}{\pi} \int_0^\pi d\phi \frac{\tanh(\tilde{\beta} \Lambda(\tilde{a})/2)}{\Lambda(\tilde{a}) \Lambda^2(0)} \nonumber \\
&\times & \cos\phi [(\cos\phi - \tilde{a})\cos\phi + \gamma^2 \sin^2 \phi]. 
\label{eq:Mzevol}
\end{eqnarray}
Here
\(\Lambda(x)= \left\{\gamma^2\sin^2\phi~+~[x-\cos\phi]^2\right\}^{\frac{1}{2}}\), and \(\tilde{\beta} = \beta J\).
The correlations  
are given as follows:
\[
T^{xx} =G(-1), \quad T^{yy} = G(1),\quad T^{zz}= [M^z]^2-G(1)G(-1)
 \]
where
$G(R)$ (for $R =\pm 1$)
is given by
\begin{eqnarray}
G(R) &=& \frac{1}{\pi}\int^\pi_0d\phi \frac{\tanh(\beta \Lambda(\tilde{a})/2)} {\Lambda(\tilde{a})\Lambda^2(0)} (\gamma \sin(\phi R)\sin \phi - \cos^2\phi) \nonumber\\
& &\times (\gamma^2 \sin^2\phi +(\cos\phi-\tilde{a})\cos\phi).
\end{eqnarray}
Moreover,  \(T^{xy} = T^{yx}=0\).
 As  \(\tilde{a} \rightarrow \infty\),
\(G(R) \rightarrow 0\) for  \(R = \pm 1\) for all \(\gamma\) and 
\(\tilde{\beta}\). Therefore, also \(T_{xx} = T_{yy}=0\), while \(T_{zz} \) 
reduces to \(4 [M^z]^2\). And 
\[M^z \rightarrow \frac{1}{1+ |\gamma|}\]
 as  \(\tilde{a} \rightarrow \infty\).  
Therefore for high fields, the two-site reduced density matrix turns out to be of the form \(\rho_A\otimes \rho_B\), with 
\(\rho_A = \rho_B = \frac{1}{2}(I + M^z \sigma^z)\), 
and hence all quantum correlation measures  of the evolved state vanish. Vanishing of quantum correlation measures is also true for the thermal equilibrium state 
for \(T \rightarrow \infty\), implying an ergodic nature for all the measures of quantum correlations. 
It also shows that \(M^z\) is  nonergodic for any choice of \(\gamma\), if one chooses \(h(t) =0\) for \(t>0\).\hfill \(\square\)

\begin{figure}[h]%
\resizebox{\columnwidth}{!}{
\includegraphics{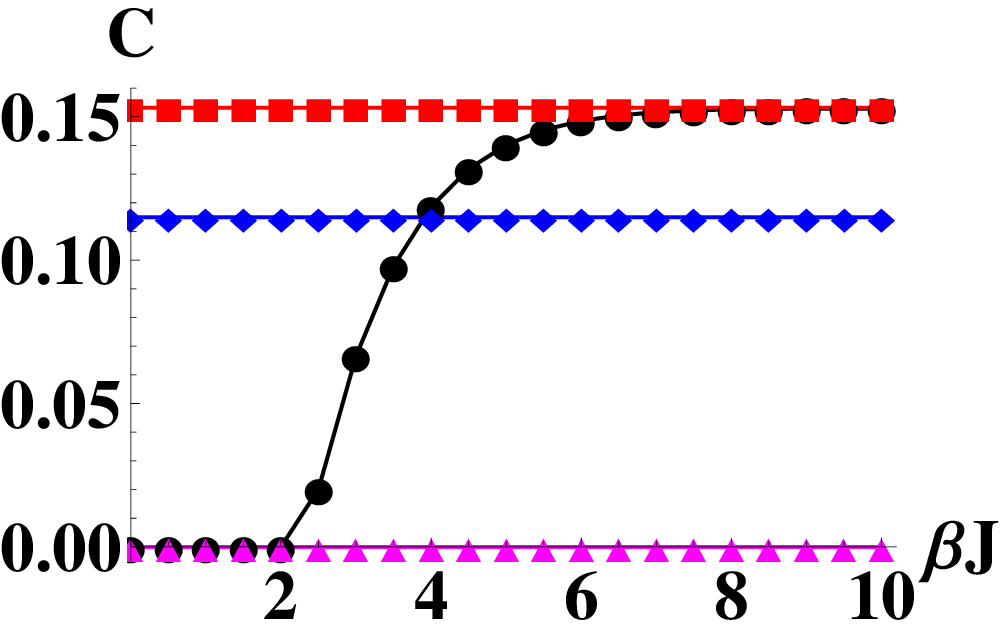}
\includegraphics{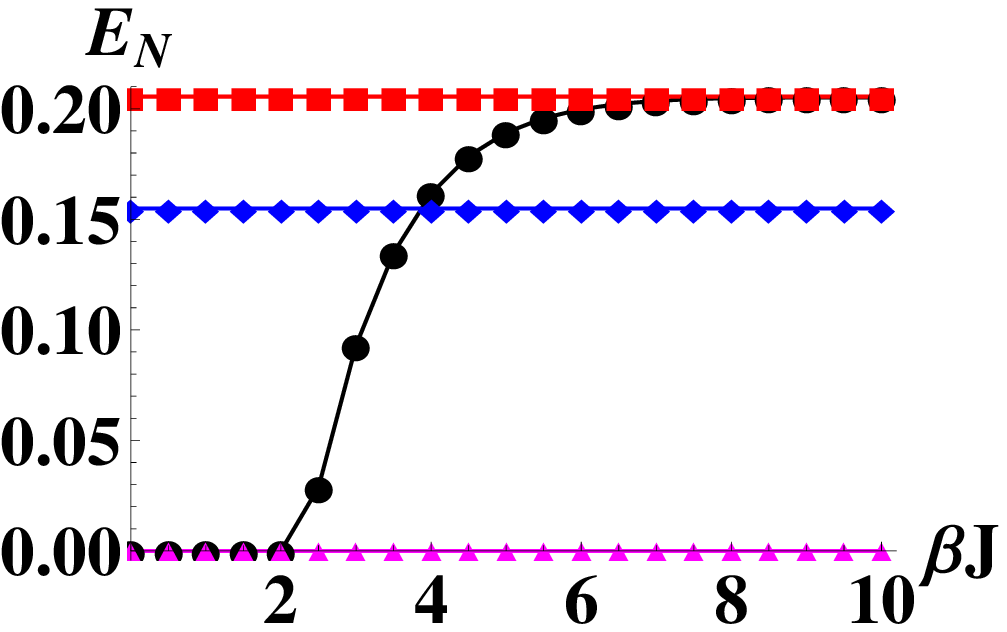}}
\resizebox{\columnwidth}{!}{
\includegraphics{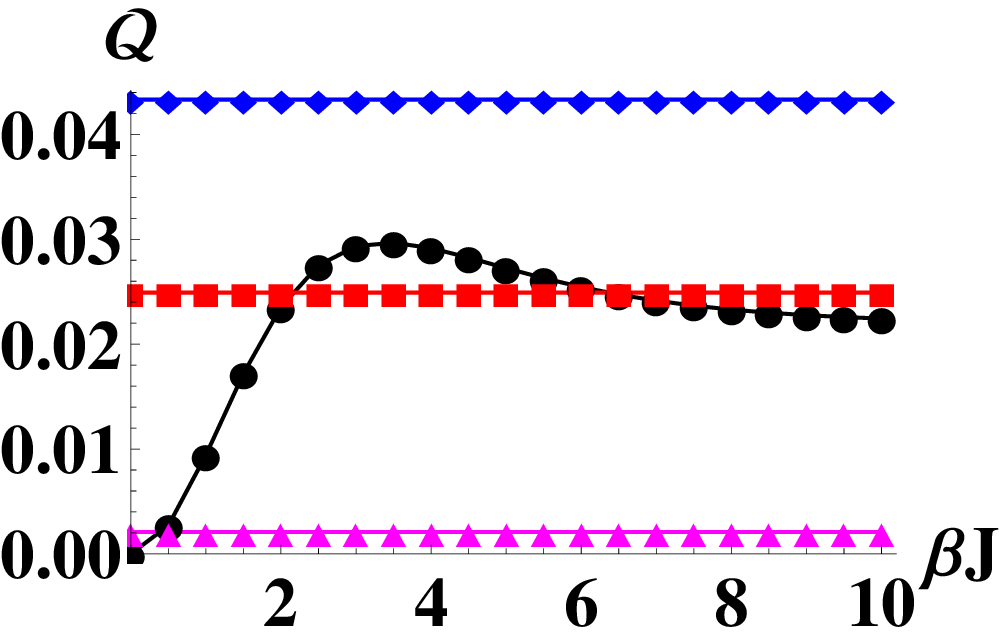}
\includegraphics{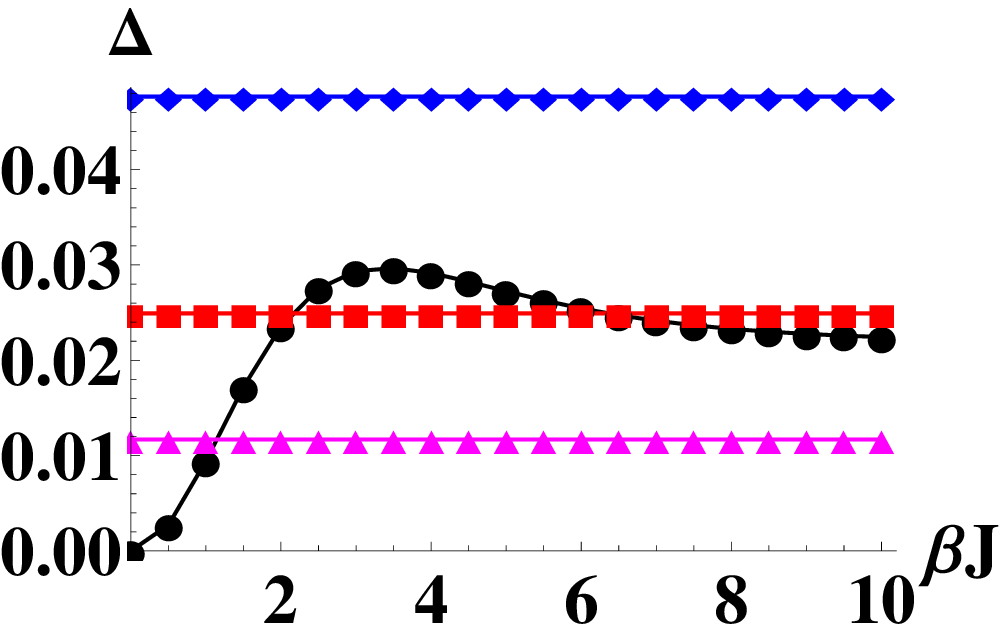}}
\caption{(Color online.) Ergodic entanglement measures versus nonergodic information-theoretic measures. 
Entanglement measures (concurrence and logarithmic negativity, in ebits) and information-theoretic quantum correlation measures (quantum discord and quantum work-deficit, in bits) are plotted on the vertical axes against the dimensionless 
parameter \(\tilde{\beta}\), for the infinite quantum transverse $XY$ spin chain. 
Concurrence ($C$, top-left) and logarithmic negativity ($E_N$, top-right) are ergodic for any value of initial transverse field $a$. In contrast, information-theoretic quantum correlations -- quantum discord ($Q$, bottom-left) and quantum work-deficit ($\triangle$, bottom-left) -- are nonergodic for moderate values of the applied transverse field.
The curves with black circles are for the canonical equilibrium state at large time, as functions of $\tilde{\beta}$. We have plotted the values of the corresponding physical quantity, for the time-evolved state with the initial state of the evolution being the canonical state at \(\tilde{\beta}=20\), as horizontal lines for different values of the applied transverse field. Precisely, the applied fields used  for the horizontal lines are  $a/J=0.2$ (red squares), $a/J=0.6$ (blue diamonds), and $a/J=2$ (pink triangles). Note that \(a/J\) is a dimensionless quantity. Here $\gamma=0.5$.}
\label{fig:1dinfall}
\end{figure}

\subsubsection*{Ergodicity of entanglement measures} Combining Propositions 1 and 2, we find that 
all quantum correlation measures behave similarly for low and high fields. The situation is drastically different for moderate fields. We begin by demonstrating that two entanglement measures, viz., concurrence and logarithmic negativity, exhibit ergodic behavior, also for moderate fields.
Let us begin by considering the behavior for concurrence 
(see Fig. \ref{fig:1dinfall} (top-left)). The computations are performed by using Jordan-Wigner transformations.
For the purpose of presentation in all the figures in this paper, we choose the anisotropy \(\gamma\) to have the value \(\frac{1}{2}\), and the 
temperature of the initial (canonical) state of the time-evolution to be such that \(\tilde{\beta} = 20\). These choices are of no particular 
significance, and all the results obtained have been confirmed to be qualitatively similar for all nonzero values of \(\gamma\) and for all low initial 
temperatures. A nonzero anisotropy is necessary for the change in the transverse field to have any nontrivial effect on the evolution, and 
high temperature canonical states are known to have zero or near-zero quantum correlations. 
For the evolved state, the time-averaged values of concurrence, at large times, for the exemplary values of $a/J$=0.2, 0.6, and 2.0 are 0.153, 0.1149, and 0 
respectively. These are depicted in the figure (Fig. \ref{fig:1dinfall} (top-left)) as horizontal lines at the corresponding heights.  
The concurrence of the canonical equilibrium state 
starts off from zero for  $\tilde{\beta}=0$
and converges to 0.153 at low temperatures. Hence, there always exists a temperature for which the corresponding equilibrium value matches the time-averaged 
value in the evolved state, up to the 3rd significant digits.   The feature is shared by other values of \(a/J\) as well. 
Therefore we conclude that the concurrence is ergodic for the infinite 1D $XY$ spin model. For low values of the initial field, the concurrence may only be canonically ergodic. 
It is clear from  Fig. \ref{fig:1dinfall} (top-right) that, along with the concurrence, the logarithmic negativity, another 
quantum correlation measure defined within the entanglement-separability paradigm, 
 is also ergodic  (or at least canonically so) in this model. 
In Ref. \cite{amaderergo1}, the question of ergodicity of logarithmic negativity
 was addressed for the 1D XY spin model by requiring that  the energy of   the equilibrium state and the dynamically evolved state is the same. 
This, however, is a stringent condition, and is not usually considered to be necessary 
for ergodicity considerations (see e.g. \cite{Barouch1075, Barouch786}). Rather surprisingly, we find that by removing this energy constraint, 
entanglement-like quantities (including logarithmic negativity) become ergodic in this model.

\subsubsection*{Nonergodicity of information-theoretic measures} 
Measures of quantum correlation defined from an information-theoretic perspective exhibit a behavior that is opposite to that of entanglement measures, for moderate fields.
Let us first consider the case of QD. For low values of the applied transverse field, the QD remains ergodic (see Fig. \ref{fig:1dinfall} (bottom-left)).
This is just like for entanglement measures and also follow directly from Prop. 1. However as the applied transverse field is increased to moderate values, 
there is no temperature for which the time-averaged value of QD 
matches with that of the canonical equilibrium value -- QD is strongly nonergodic in the vicinity of \(\tilde{a} = 0.6\). Again the computations are performed by using successive Jordan-Wigner and Fourier transformations.
%
For even higher magnetic fields -- 
e.g. for \(a/J = 2.0\), -- 
QD attains ergodicity only if we allow temperatures much higher -- precisely
two orders of magnitude higher --
than
the initial temperature, and hence we still term the situation as nonergodic, although not strongly so.

Surprisingly, we have also observed that QD shows nonergodicity even when its constituent quantities, \(\mathcal{I}\) and \(\mathcal{J}\), are both ergodic. 
This arises due to the fact that \(\mathcal{I}\) and \(\mathcal{J}\) become ergodic at different temperature ranges, and so the ergodicity of QD does not remain a linear function of the ergodicities of \(\mathcal{I}\) and \(\mathcal{J}\). 
 QD must be ergodic for $a\rightarrow \infty$, follows from Prop. 2.
Very  similar features are observed in the other information-theoretic quantum correlation measure, viz., quantum work-deficit 
(see Fig. \ref{fig:1dinfall} (bottom-right)).

\begin{figure}[h]%
\resizebox{\columnwidth}{!}{
\includegraphics{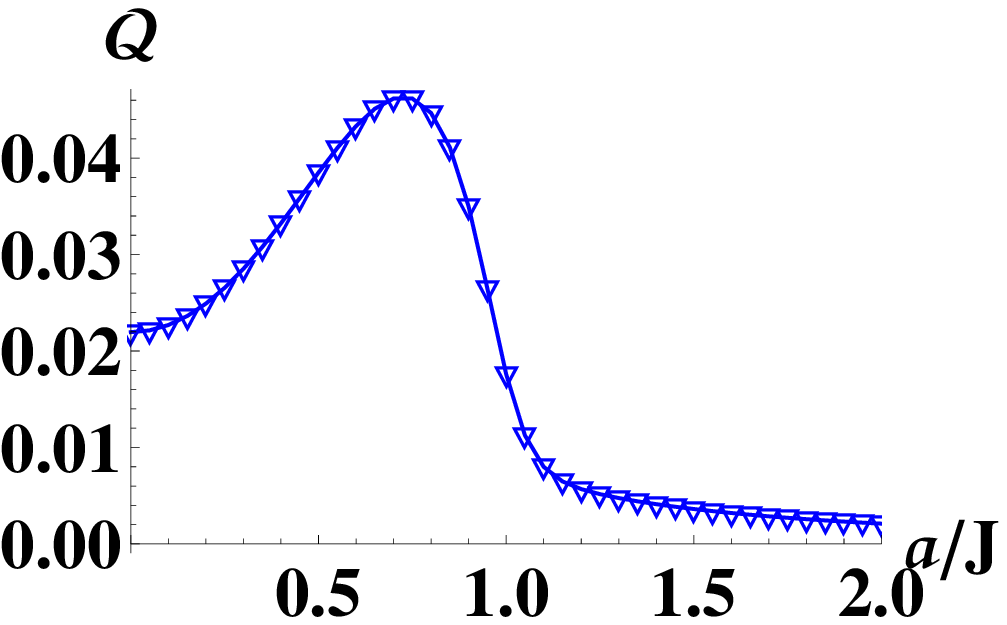}
\includegraphics{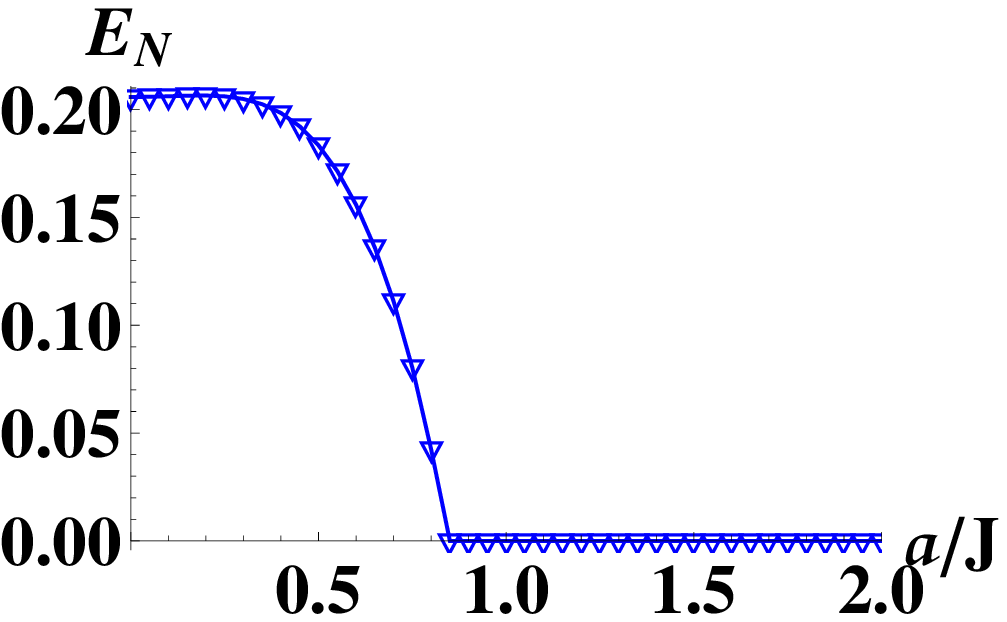}}
\caption{(Color online.) Quantum discord versus logarithmic negativity at large time. Quantum discord (in bits, left figure) and logarithmic negativity (in ebits, right figure)  of the time-evolved state at $t \rightarrow \infty$ are plotted on the vertical axis, for the infinite quantum transverse $XY$ spin chain is plotted against the dimensionless transverse magnetic field parameter $a/J$, on the horizontal axis. 
The temperature of the initial canonical state of the evolution is taken such that \(\tilde{\beta}=20\), and $\gamma=0.5$.}
\label{fig:1dinfde}
\end{figure}

\subsubsection*{Discussion} In an attempt towards  understanding the opposite
behavior of the quantum correlations in the two prominent paradigms, let us begin by contrasting QD against logarithmic negativity for 
the evolved state, with respect to the initial transverse field at large time  (see Fig. \ref{fig:1dinfde}). 
We find that while logarithmic negativity is monotonic (decreasing) for the whole range of \(\tilde{a}\),   QD is increasing for  low values of magnetic field and 
decreasing for high values. 
This different behavior of the two measures 
%
hints at 
the different statistical-mechanical behavior for QD and logarithmic negativity. 

Another way to interpret the opposite behavior observed is the following. Classical correlators, given by $T_{ij}= \left\langle\sigma_i\otimes\sigma_j \right\rangle$, 
of the states of a many-body quantum system has long been known in the literature of condensed matter \cite{many-body-physics} and 
Bell inequalities \cite{BellBell}, among others, as important characterizing quantities of the physical system. 
Quantum correlations have recently been identified to be interesting quantities of 
many-body systems
\cite{Lewenstein07, Amico08, Bloch08}. 
A typical multisite quantum state contains both quantum and classical correlations. The measures of entanglement 
strives to 
filter out only the quantum part of the correlations in a way that is very stringent, as compared to  the information-theoretic measures
of quantum correlations. Indeed, even separable states have nonzero quantum discord  \cite{discord1} and quantum work-deficit \cite{deficit}. 
Therefore, it is reasonable that the information-theoretic measures of quantum correlations will behave more closely like the 
classical correlators, while measures defined within the entanglement-separability paradigm will behave oppositely thereof. 
The classical correlators 
in these systems have long been known to be generally nonergodic \cite{Mazur69,Perk77,Barouch1075,Barouch786} (cf. \cite{amaderergo2}).
They are ergodic for low transverse fields, while becoming nonergodic for moderate fields -- exactly the feature obtained for 
quantum discord and quantum work-deficit, while exactly the opposite to that obtained for concurrence and logarithmic negativity.


\emph{Magnetic field regions.--}  It is clear from Fig. 2 that whenever the applied field strength $a/J$ is approximately less than 0.4, all the quantum correlation measures, whether they 
belong to the entanglement-separability paradigm or the information-theoretic one, have a low derivative (slow change) with respect to the applied field. We call such regions as ``low field regions''. Similar behavior (of the measures) 
is also observed when $a/J$ is approximately above 1.0. Such regions are called the ``high field regions''. The remaining fields (on the positive field axis) fall in what we term as the ``moderate field region''. 
In this moderate field region, all the quantum correlation measures have a substantially larger derivative values (in comparison to the same in the other regions) with respect to the applied field. It may be noted here that the quantity 
$a/J$ is a dimensionless quantity, so that the field (\(a\)) regions are already defined in energy (\(J\)) units.

\section{Finite Quantum $XY$ spin chain in a transverse field}
\label{sec:finite1d}

In the previous section, we considered the infinite $XY$ spin chain in a transverse field and found that information-theoretic quantum correlation measures and those of the
entanglement-separability paradigm 
have countering statistical-mechanical behavior. It will be interesting to see whether such complementary behavior is 
sustained even in higher-dimensional systems. However, the analytical methods do not exist for a ladder or a two-dimensional model and so  we have to use numerical 
simulations to study them. The question remains whether results obtained by exact diagonalization can infer correctly the ergodic behavior  
of ladder or two-dimensions. Hence, we first consider the finite $XY$ spin chain of moderate length 
and compare their results with the infinite one.


\begin{figure}[h]%
\resizebox{\columnwidth}{!}{
\includegraphics{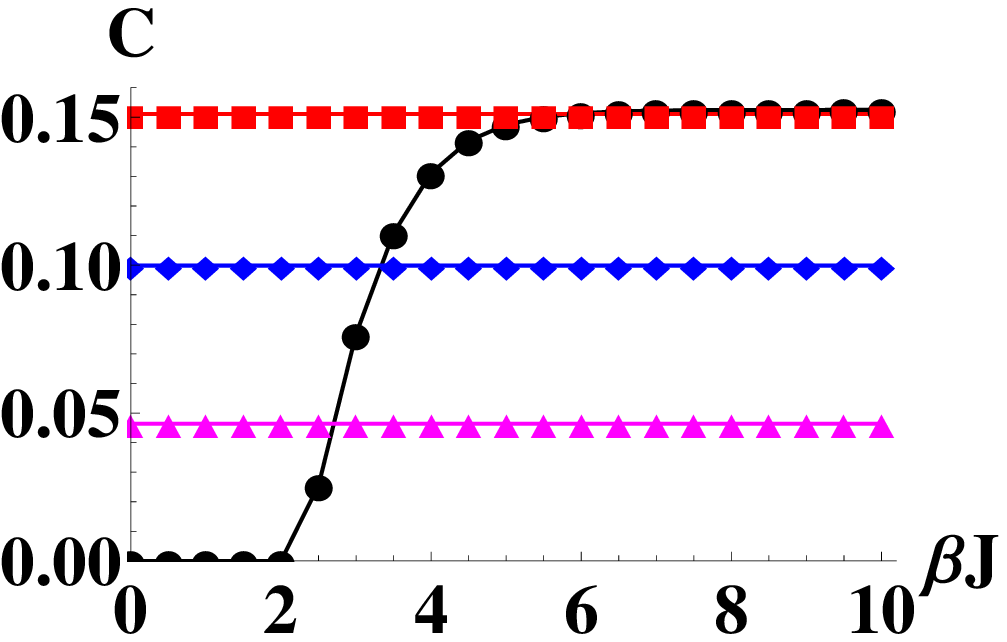}
\includegraphics{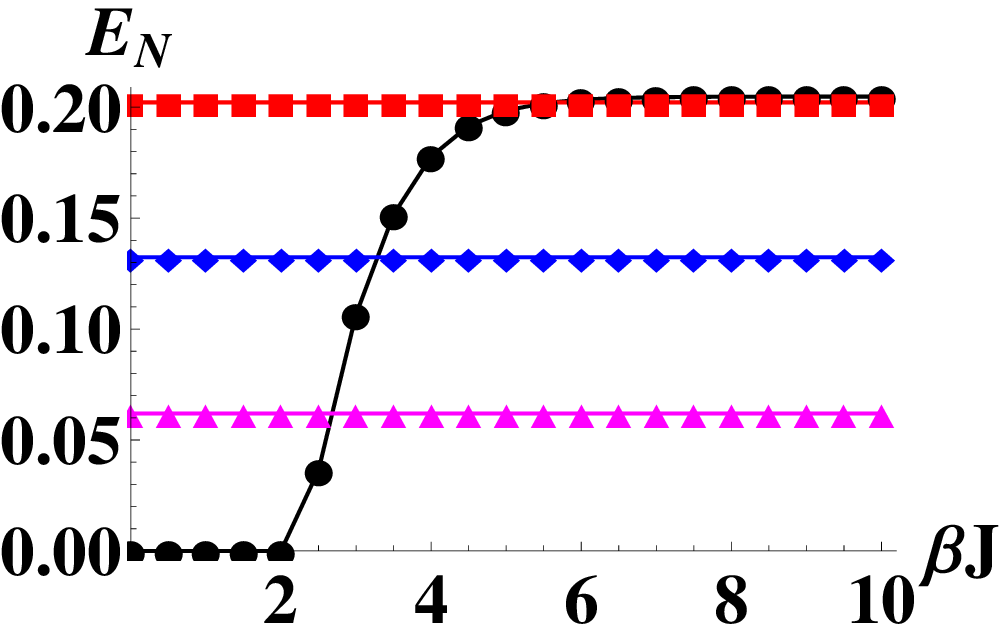}}
\resizebox{\columnwidth}{!}{
\includegraphics{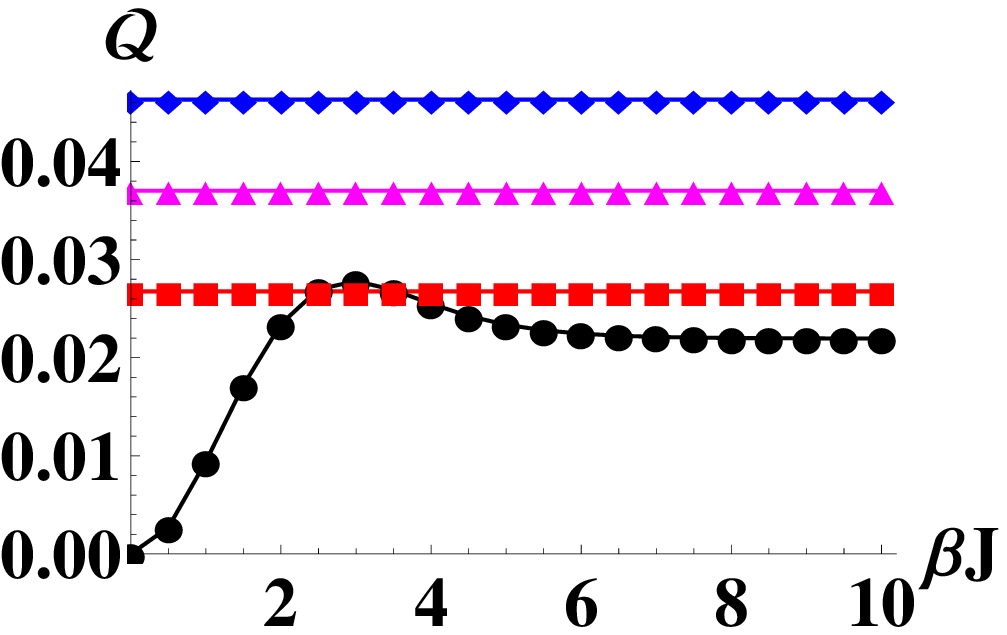}
\includegraphics{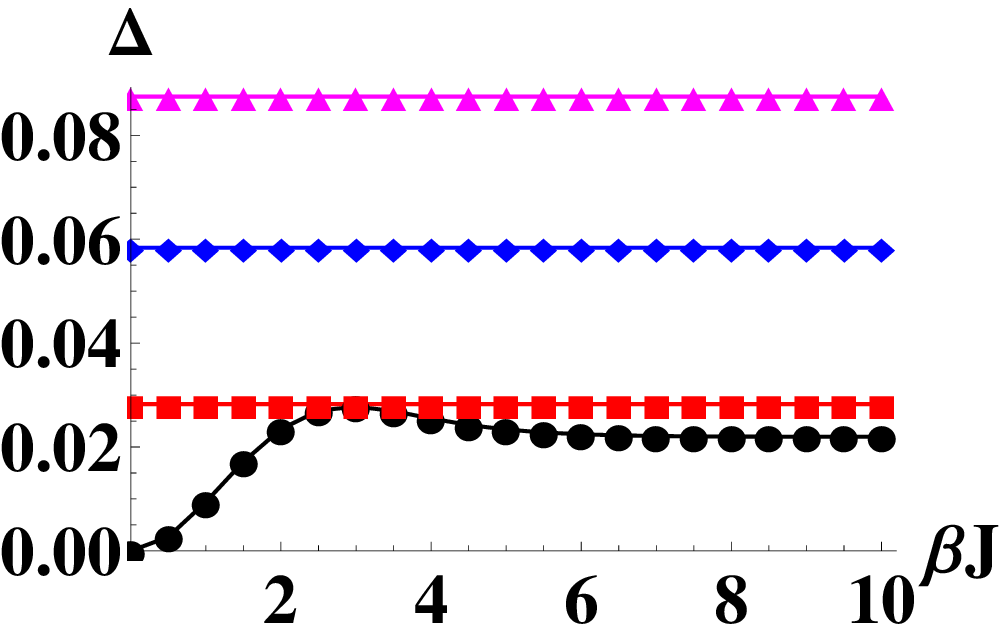}}
\caption{(Color online.) Reproducing the content of Fig. \ref{fig:1dinfall} for finite chain of 12 spins with periodic boundary conditions. All considerations remain the same as in Fig. \ref{fig:1dinfall}. 
}
\label{fig:1dn12}
\end{figure}

The Hamiltonian for a 
spin chain of 12 spins, with periodic boundary conditions,  is given by
\begin{equation}
\label{12H}
H = J \sum_{i=1}^{12} \left[(1 + \gamma) S_i^x S_{i+1}^x + (1 - \gamma) S_i^y S_{i+1}^y\right] -  
h(t) \sum_{i=1}^{12} S_i^z.
\end{equation}
Similar to the case of the infinite spin chain, an initial transverse field is applied as a disturbance 
on the initial canonical equilibrium state at $\tilde{\beta}=20$, and then we allow the system to evolve. Later, we compare the long time-average of the physical quantities 
for the evolved state and the canonical equilibrium state at different field strengths. The plots of concurrence, logarithmic negativity, quantum discord, and quantum 
work-deficit for the evolved state and the canonical equilibrium as functions of temperature are given in Fig. \ref{fig:1dn12}. Similar to the case of infinite $XY$ 
chain, we observe that entanglement-based measures, viz., concurrence and logarithmic negativity,  remain ergodic for the entire range of the applied initial transverse magnetic field. However, the information-theoretic quantum correlation measures are ergodic for low and high values of applied transverse field and nonergodic at moderate values.
To illustrate this, let us consider the case of QD, which remain ergodic for low
initial fields (Fig. \ref{fig:1dn12} (bottom-left)).
For moderate 
values of the initial field, QD becomes nonergodic, i.e, there is no temperature for which the time-averaged value of QD for the evolved state is equal to the QD of the equilibrium state. 
Hence, the statistical-mechanical behavior of the quantum correlation measures defined in the entanglement-separability paradigm (concurrence and logarithmic negativity) and that of the measures defined in the information-theoretic one (quantum discord and quantum work-deficit) for finite  1D $XY$ model 
of length $N=12$, mimics the  properties of the infinite spin chain.
This induces us to study such statistical-mechanical behavior in the case of similar Hamiltonians in
higher dimensions, where we will be forced to consider finite-sized systems, as these systems are not analytically diagonalizable.
This will be carried out in the two succeeding sections.

\section{Quantum $XY$ ladder model in a transverse field}
\label{sec:ladder}

The ladder arrangement of spins can be thought of as
an intermediate dimension between one and two dimensions. These systems are interesting  as 
they have revealed several intriguing features of both one and two dimensions. Ladder models have been successfully realized in laboratories in certain materials 
and currently available techniques in cold gas systems also offer the possibility of realizing such models \cite{ladder-reference}.
The ladder model with periodic boundary conditions and $XY$ interactions on steps as well as rungs is considered.

\begin{figure}[h]%
\resizebox{\columnwidth}{!}{
\includegraphics{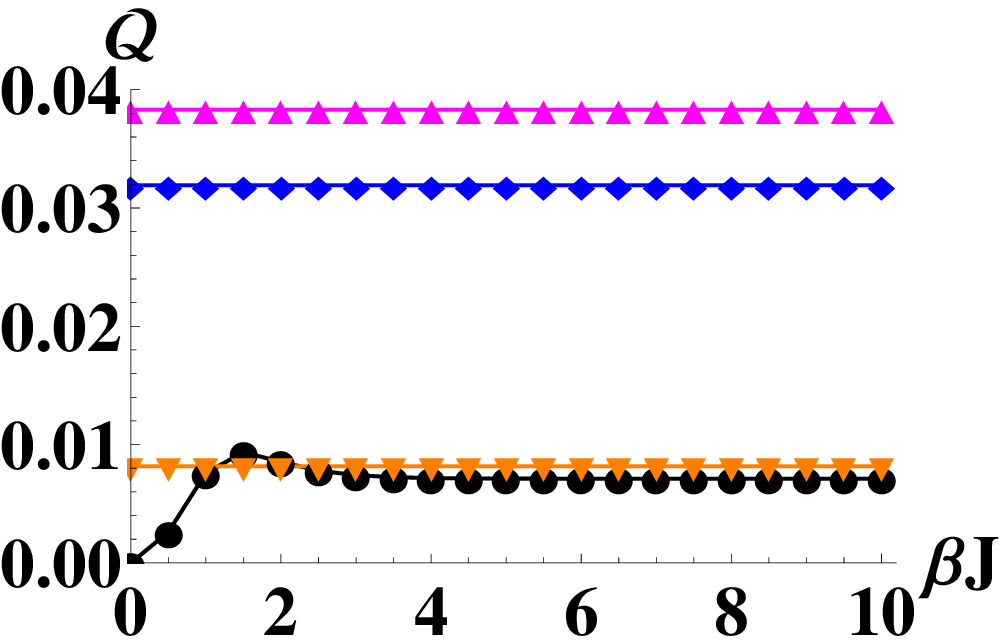}
\includegraphics{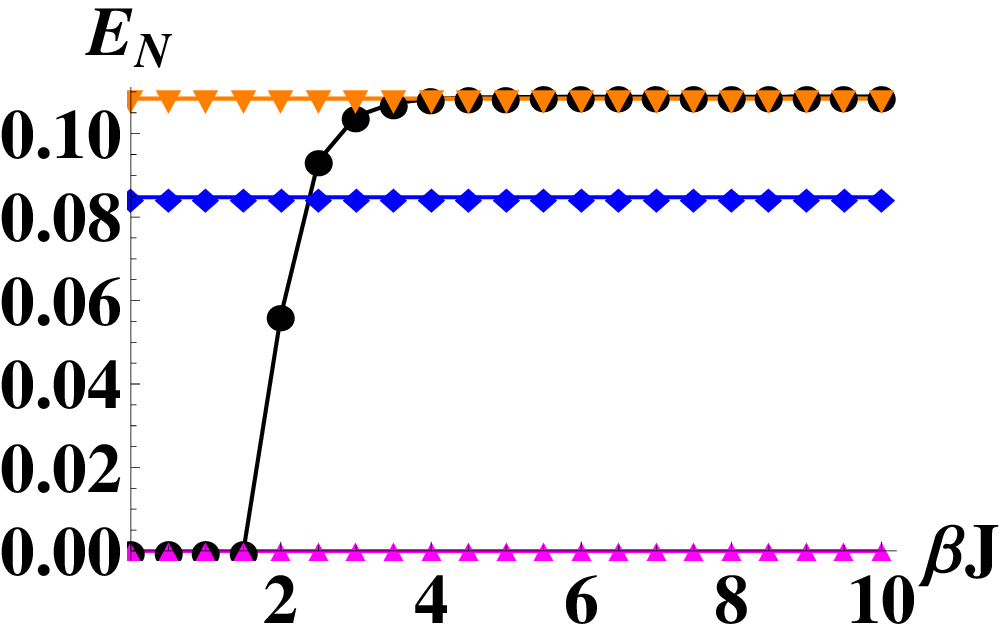}}
\caption{(Color online.) Ergodicity versus nonergodicity of quantum correlation measures on \(XY\) ladders. 
Quantum discord (left, in bits) and logarithmic negativity (right, in ebits) for a ladder of eight spins with periodic boundary condition. The considerations are the same as in Fig. \ref{fig:1dinfall}, except for the orange upside-down triangles which are for $a/J=0.1$. 
}
\label{fig:ladn8}
\end{figure}

Again the system is evolved from the canonical equilibrium state with an initial temperature of $\tilde{\beta}=20$, after applying a kick in the transverse magnetic field. As in the case of the infinite and finite spin chains, 
for moderately high fields, QD and quantum work-deficit are 
nonergodic, while nearest-neighbor entanglement (concurrence as well as logarithmic negativity) of steps as well as rungs remains ergodic. From Fig. \ref{fig:ladn8} (right), it is clear that the evolved state logarithmic negativity, plotted as horizontal lines corresponding to transverse field strengths of $a/J$=0.1, 0.6, and 2.0, will definitely cross the logarithmic negativity of the canonical equilibrium state for some values of the temperature.
Such crossing of the equilibrium and evolved state entanglement curves does not happen 
for QD for initial fields \(a/J > 0.15\) (Fig.   \ref{fig:ladn8} (left)).
Thus the statistical behavior of nearest-neighbor entanglement-based measures and information-theoretic ones, qualitatively matches with those of the $XY$ spin chain: entanglement measures are ergodic while information-theoretic 
ones are not so, for moderate field strengths.

\section{Transverse 2D $XY$ spin chain}
\label{sec:2d}
The two-dimensional quantum \(XY\) model with a transverse magnetic field plays an important role in both many-body physics and quantum information. In particular, the dynamics of this model in the case of unit anisotropy is used for realizing the measurement-based quantum computation \cite{NP-Briegel}.
Here we consider a two-dimensional array of 12 spin-$\frac{1}{2}$ particles arranged on a square lattice, and interacting via nearest-neighbor \(XY\) interactions with periodic boundary conditions. 
The arrangement therefore forms a torus. 

\begin{figure}[h]%
\resizebox{\columnwidth}{!}{
\includegraphics{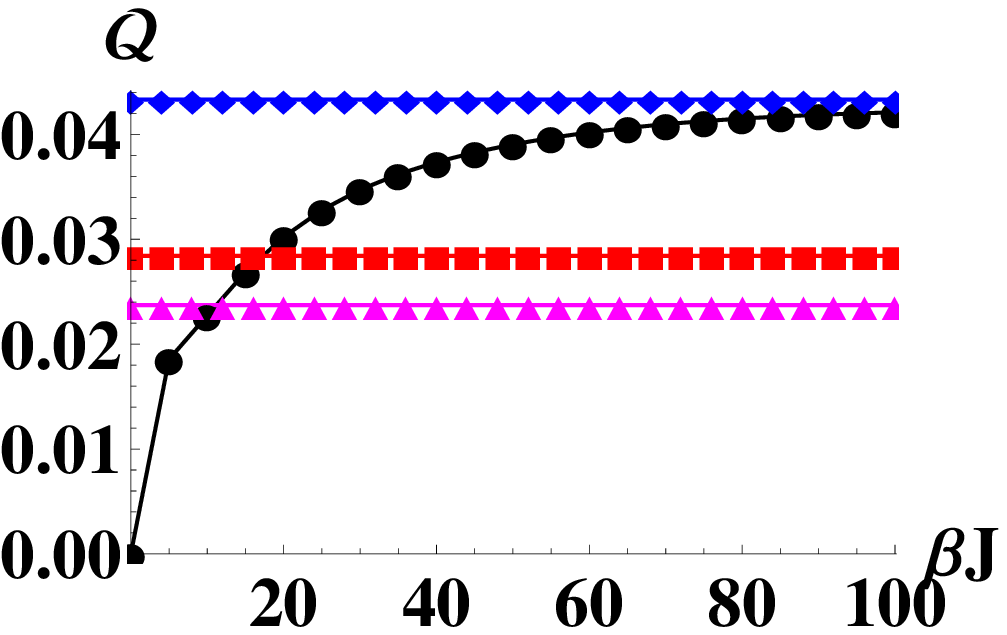}
\includegraphics{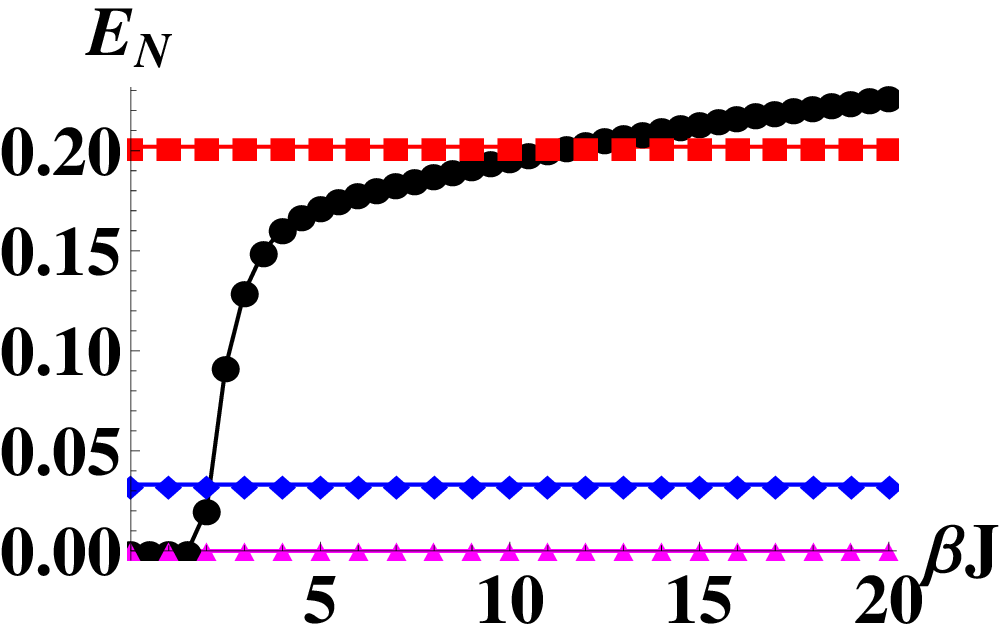}}
\caption{(Color online.) Ergodicity versus nonergodicity on 2D \(XY\) models. Quantum discord (left, in bits) and logarithmic negativity (right, in ebits) for a 2D array of 12 spins with 
periodic boundary condition. 
All considerations are the same as in Fig. \ref{fig:1dinfall}. 
}
\label{fig:2dn12}
\end{figure}

For various transverse field strengths, viz., \( a/J=0.2, 0.6,\) and \( 2.0\), we find that the entanglement-based measures 
show ergodicity also in this case. 
However, we find that for $a/J=0.6$, the time-averaged QD of the evolved state has the value 0.0433 for an initial temperature, corresponding to $\tilde{\beta}=20$, while the QD of the canonical equilibrium state  converges to 0.0424, and hence 
leads to  nonergodicity. The same feature is shared by quantum work-deficit.
In contrast, entanglement-based measures remain ergodic for all values of the transverse field, including $a/J=0.6$.  

One may expect that the ergodic behavior of the quantum correlations in the 2D model will be  different from that in the 1D model. However, simulations in  finite 2D systems show a
qualitatively similar behaviour as in 1D. 
However, some classical correlations (of the form $T^{ij}$) turn out to be ergodic in 2D, while they are not so in 1D \cite{amaderergo2}.

\section{conclusion}
\label{summary}

We have considered the statistical-mechanical properties of quantum correlations of the transverse quantum \(XY\) model with nearest-neighbor interactions in low dimensions with periodic boundary conditions.
We have shown that any quantum correlation measure of a Hamiltonian with a time-independent interaction and a non-commuting time dependent field is 
ergodic for low fields, irrespective of the other parameters of the Hamiltonian. 
Moreover, for the infinite $XY$ spin chain, with a high initial transverse magnetic field, we have analytically demonstrated that any quantum correlation measure is again ergodic.
Furthermore, we found that the quantum correlation measures defined in the entanglement-separability paradigm, 
equilibrate with time, and hence remain ergodic, while the information-theoretic quantum correlation measures  
can be nonergodic for moderate initial fields. 
The thesis is true for the infinite chain, which is exactly diagonalizable. 
Low-dimensional finite systems, that cannot be analytically diagonalized -- ladder and two-dimensional models with periodic boundary conditions -- mimics the one-dimensional infinite one. 
We therefore obtain a distinct partition that divides the world of quantum correlation measures -- ones that are ergodic as opposed to the ones that are not so, in low-dimensional quantum transverse $XY$ spin-$\frac{1}{2}$ systems. 
More interestingly, the partition coincides with that which divides them into entanglement-based quantum correlation measures and information-theoretic ones. 

\begin{acknowledgments}

RP acknowledges an INSPIRE-faculty position at the Harish-Chandra Research Institute (HRI) from the Department of Science and Technology, Government of India.
We acknowledge computations performed at the cluster computing facility in HRI. 
\end{acknowledgments}

\end{document}